\documentclass[twocolumn,prl,showpacs,aps,superscriptaddress]{revtex4}
\usepackage{epsfig}
\usepackage{dcolumn}
\usepackage{color}

\begin{document}

\title{Break-up of long-range coherence due to phase fluctuations in ultrathin superconducting NbN films}

\author{Y. Noat}
\author{T. Cren}
\author{C. Brun}
\author{F. Debontridder}
\author{V. Cherkez}
\affiliation{Institut des Nanosciences de Paris, CNRS-UMR 7588, Universit\'e Pierre et Marie Curie-Paris 6 UPMC, 4 place Jussieu, 75252, Paris, France}

\author{K. Ilin}
\author{M. Siegel}
\affiliation{Institute of Micro- und Nano-electronic Systems, Karlsruhe Institute of Technology, Hertzstrasse 16, D-76187 Karlsruhe, Germany}

\author{A. Semenov}
\author{H.-W. H\"ubers}
\affiliation{DLR Institute of Planetary Research, Rutherfordstrasse 2, 12489 Berlin, Germany}

\author{D. Roditchev}
\affiliation{Institut des Nanosciences de Paris, CNRS-UMR 7588, Universit\'e Pierre et Marie Curie-Paris 6 UPMC, 4 place Jussieu, 75252, Paris, France}

\email{noat@insp.jussieu.fr}

%\date{\today}

\begin{abstract}

Using scanning tunneling spectroscopy (STS), we address the problem of the superconductor-insulator phase transition (SIT) in homogeneously disordered ultrathin (2-15~nm) films of NbN. Samples thicker than 8~nm, for which the Ioffe-Regel parameter $k_F l \geq 5.6$, manifest a conventional superconductivity : A spatially homogeneous BCS-like gap, vanishing at the critical temperature, and a vortex lattice in magnetic field. Upon thickness reduction, however, while $k_F l$ lowers, the STS revealed striking deviations from the BCS scenario, among which a progressive decrease of the coherence peak height and spatial inhomogeneities. The thinnest film (2.16~nm), while not being exactly at the SIT ( $T_C \approx 0.4 T_{C-bulk}$), showed astonishingly vanishing coherence peaks and the absence of vortices. In the quasi-2D limit, such clear signatures of the loss of long-range phase coherence strongly suggest that, at the SIT the superconductivity is destroyed by phase fluctuations.

%the gap below $T_C$ develops on a spectral background which becomes more and more "V"-shape close to the localization. In addition, spatial inhomogeneities of the superconducting state and the absence of a vortex lattice were revealed. The data analysis evidenced  two distinct spectral features to exist in the tunneling spectra with different energy range: One corresponding to the superconducting condensate, and another, reflecting a non-trivial normal state on which the superconductivity sets in at $T<T_C$.

%In the thickest films (15nm and 8nm), the spectra are quite homogeneous and well described by the Bardeen-Cooper-Schrieffer theory of superconductivity. The situation is radically different in the two thinnest films (2.33nm and 2.17nm), where the spectra deviate strongly from BCS and are inhomogeneous on the nanometer scale.

\end{abstract}


\maketitle

Despite numerous theoretical and experimental works \cite{Gantmakher2010}, the understanding of the phase transition from a superconducting (SC) to an insulating state in ultrathin films remains a very challenging problem. The structural properties of the films are known to play a key-role in determining the low-temperature electronic properties. Disordered SC thin films undergoing a superconductor-insulator transition (SIT) can be divided into two groups \cite{Goldman} : i) Granular systems, consisting of coupled SC grains, described at low temperature as an array of Josephson junctions \cite{Orr1985}; ii) Systems qualified as "homogeneous" where both amplitude and phase variations of the order parameter are believed to play an important role \cite{Graybeal}. In contrast, in perfectly crystallized systems the SIT was not observed down to a single atomic monolayer \cite{Zhang2010}. %These two groups present significant differences in the electronic properties at low temperature. For example while in (i) the SC gap ($\Delta$) and critical temperature ($T_c$) remain close to the bulk values on approaching the SIT \cite{Orr1985}, in (ii) $\Delta$ and $T_c$ gradually diminish \cite{Graybeal} and eventually tend to vanish at the SIT \cite{Valles}}.

A way to get a new insight in the microscopic processes occurring at the SIT is to study the local properties of the SC condensate evolving in a disordered potential. Recently two scanning tunneling microscopy and spectroscopy (STM/STS) experiments addressed this question bringing new interesting results and leading to various successive interpretations \cite{Sacepe_TiN,Sacepe_InOx}. It is worth noting that the SIT problem has also important application issues, specifically in the field of single photon detectors \cite{Goltsman} in which ultrathin SC films (usually NbN or NbTiN) are used. While the microscopic picture of the photon-to-condensate excitation conversion does not imply any thickness dependence, it has recently been discovered that the maximum detection efficiency occurs for film thicknesses close to the SIT \cite{Hoferr}. Thus, the detailed study at the nanometer scale of such SC films would be of help in the understanding of the underlying microscopic processes of the photon absorption by \emph{real} SC devices. %In TiN films (3.6-5~nm) slight inhomogeneities of the SC gap ($\sigma/\overline{\Delta} \sim 4$ to $6\%$) were observed together with a deep pseudogap background below and above $T_c$ \cite{Sacepe_TiN}. While the authors first concluded that their results were consistent with various theoretical models, it was further argued that superconducting fluctuations, which effect was fitted in a small temperature range ($1.1-2.7~T_c$ ), could explain the pseudogap state. However this pseudogap was observed in a much larger temperature range (up to $14~T_c$ for the most disordered sample) which suggests other possible mechanisms. In InO films (15 and 30~nm) deep pseudogap was also observed above $T_c$ together with the disappearance of coherence peaks at $T_c$ \cite{Sacepe_InOx}. To explain the InO results and to reinterpret the TiN ones, another mechanism based on localization of preformed Cooper pairs was proposed. It is seen that the successive interpretations differ. Additionally former results on InO films \cite{Hebard} together with their large thickness seem to suggest that InO behaves as a granular system while it is presented in \cite{Sacepe_InOx} as "homogeneous".

In this Letter, we address the SIT problem locally in a direct STS experiment by varying the thickness of ultrathin NbN films from 15 down to 2.16~nm, the corresponding $T_C$ varying from $15.0$ to $6.7$~K, while the Ioffe-Regel parameter $k_F l$ decreases from $5.7$ to below $2.6$. Our STS data reveal profound changes in the local behavior of the SC films as the SIT is approached. The amplitude of the coherence peaks diminishes to almost vanish when $T_C \approx 0.4 T_{C-bulk}$, which is not in the very vicinity of the insulating transition. This feature is accompanied by a significant broadening of the quasiparticle density of states (DOS) and by the development of a correlated electronic background evidenced by a characteristic V-shape, extending to voltages 5-10~mV significantly larger than $\Delta/e$. In addition, a \emph{shallow} pseudogap regime is observed in a very narrow temperature window just above $T_C$. Interestingly, our results on 2D-superconductors present many similarities with the results reported for 3D-NbN films of comparable $T_C$ and $k_F l$ \cite{Chockalingam} but are in striking contrast with the recently reported large pseudogap regimes observed in TiN and InO \cite{Sacepe_TiN,Sacepe_InOx}. We ascribe these differences to a more homogeneous crystalline structure of our NbN films and to film conductances farther away from the critical one. Finally, while the vortex lattice is observed in thicker films subject to magnetic field, the STS contrast decreases when reducing the film thickness; strikingly, the 2.16~nm film with $T_C \approx 0.4 T_{C-bulk}$ does not show any evidence of the vortex state, emphasizing thereby the loss of long-range SC phase coherence. Our results on 2D-homogeneously disordered superconductors therefore strongly support that phase fluctuations drive the insulating transition and have already profound effects before the SIT.

%\begin{figure}
%\begin{center}
%\includegraphics[width=7.5cm]{Fig1.eps}
%\includegraphics[width=6cm,angle=90]{BCS.eps}
%\end{center}
\begin{figure}[t]
\hspace{-0.3 cm} \epsfxsize = 8.7 cm \epsfbox{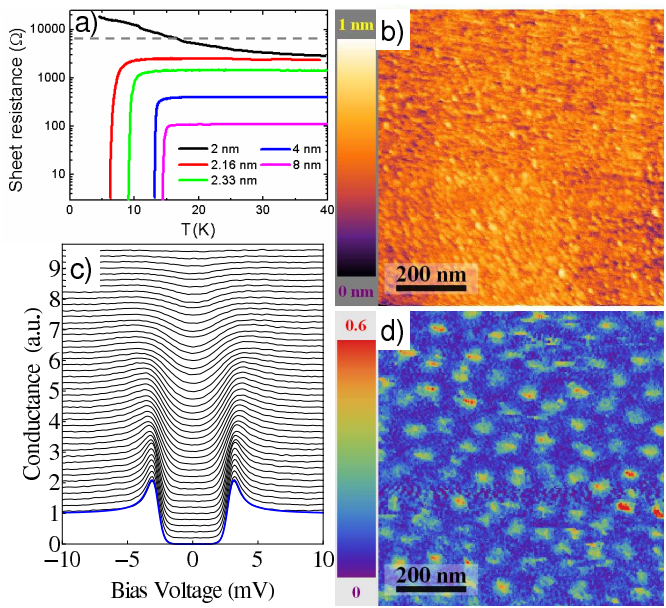}
\caption{(Color online)
a) Square resistivity vs temperature dependence of the studied samples; the dashed line indicate $h/4e^2$ the quantum resistance for pairs; $T_C$ decreases with film thickness ($T_C^{15nm}=15.0$~K (not shown), $T_C^{8nm}=14.5$~K, $T_C^{4nm}=13.3$~K, $T_C^{2.33nm}=9.4$~K, $T_C^{2.16nm}=6.7$~K); We defined $T_C$ by the temperature reached when the resistivity equals $10\%$ of its 40~K value. b) Typical topographic STM image of the surface (15~nm-thick film, $T=4.2$~K, image size $1 \mu m \times 1 \mu m $); c) The evolution of the tunneling conductance spectra (thin black lines) with temperature follows a conventional BCS behavior in the 15~nm sample: a SC gap around zero-bias, $\Delta (0)=2.85$~meV, which vanishes at the critical temperature of the film, with $\frac{2 \Delta (0)}{kT_C}=4.4$; the bottom thick blue line is a BCS fit at 2.3~K with no additional broadening parameter. d) $dI/dV$ conductance map at $V=0$ in magnetic field ($B=1$~T), revealing a disordered vortex lattice in the 15~nm film.
} \label{Fig1}
\end{figure}

In order to achieve as realistic conditions as possible, ultrathin NbN films used in real single photon detectors \cite{Hoferr} were probed; for every thickness the films were grown using identical growth conditions on crystalline sapphire substrates by DC reactive magnetron sputtering of a pure Nb target in an Ar+N$_2$ gas mixture. Their structural, optical, electronic and SC properties were extensively studied and are thus well established \cite{Semenov}; 12~nm thick films have a $ T_C$ about 15~K and $k_F l \approx 5.7$; these parameters hardly change for 6~nm thick films for which $ T_C$ is about 14~K and $k_F l \approx 5.6$; reducing further the thickness by two, 3.2~nm, $ T_C$ is reduced to 10.7~K and $k_F l \approx 2.6$; the structural characterization shows that such NbN films are covered by a 0.5-1~nm thin natural passivation layer \cite{Semenov}. Estimation of the Ginzburg-Landau coherence lengths reveals values between 4 and 6 nm; thus films having thicknesses below about 5~nm can be considered as true 2D SC, \emph{the passivation layer having to be considered as belonging to the system itself}. For thicker films, we see from the $T_C$ values very close to bulk $T_C \approx 15.5$~K, that the passivation layer induces a negligible inverse proximity effect on SC.

$R_{sq}(T)$, the square resistivity vs temperature dependence of the studied films, is presented in Fig.1a. It shows a typical SIT behavior of non-granular films : $R_{sq}(T)$ keeps a very smooth dependence on approaching $T_C$; $T_C$ monotonically lowers as the film thickness decreases. Being weak for thicker films (8~nm and more), this effect becomes significant for thinner films, reflecting the rapid suppression of $T_C$ on approaching the SIT. %For the 2.33~nm-thick film and specifically, in the case of the 2.16~nm-thick one, a characteristic insulating trend in $R(T)$ is observed above $T_C$.
Among the studied samples, the 2.16~nm one is the closest to the SIT, "sitting" on its SC side, and having $T_C \approx 0.4 T_{C-bulk}$; yet thinner films (2.0~nm) show an insulating behavior at low temperature.

%\begin{figure}
%\begin{center}
%\includegraphics[width=7.5cm]{Fig2.eps}
%\end{center}
\begin{figure}[h]
\hspace{-0.3 cm} \epsfxsize = 8.5 cm \epsfbox{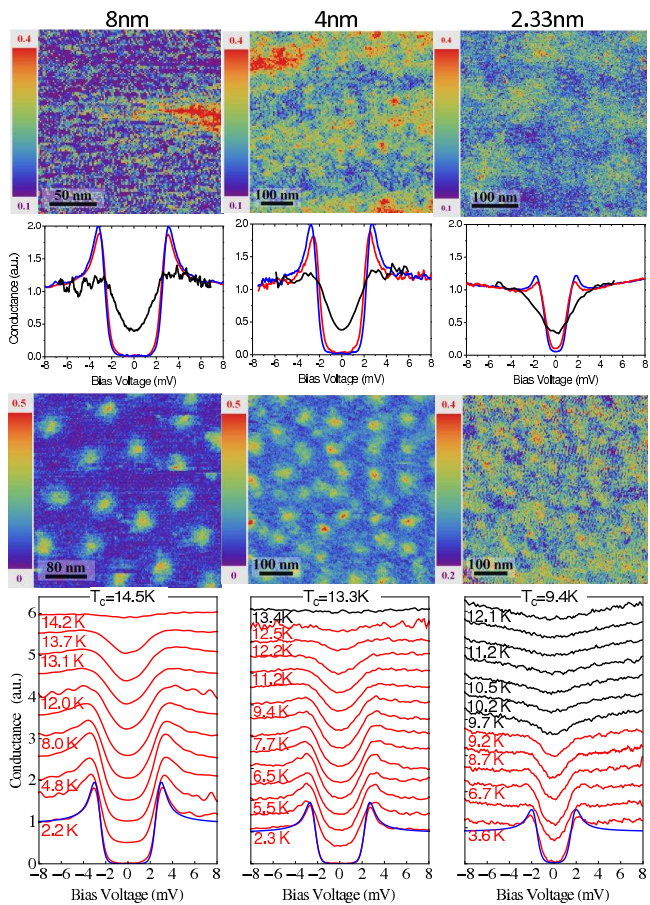}
\caption{(Color online)
From left to right column: tunneling characteristics of 8, 4 and 2.33~nm thick films. From top to bottom: the first row shows normalized STS conductance maps representing the color-coded spatial variations of the $dI/dV$ signal measured at $2.3$~K at the gap onset (for $V=1.7$~mV, $V=1.7$~mV and $V=1.1$~mV); second row: normalized individual tunneling conductance spectra representative of the spatial variations observed among the local conductance spectra (red and blue curves) at $2.3$~K; the black curves show typical spectra measured at $4.0$~K in magnetic field ($B=1$~T, $B=1$~T and $B=2$~T) at the vortex centers; third row: $dI/dV$ conductance maps at $V=0$ showing images of the vortex lattices at $4.0$~K (for $B=1$~T, $B=1$~T and $B=2$~T); fourth row: temperature evolution of the normalized $dI/dV(V)$ tunneling spectra (vertically shifted for clarity); the temperature is indicated below the corresponding spectra. Red (black) curves correspond to the SC (normal) state. The bottom thick blue lines are BCS $dI/dV$ calculations at the indicated temperature; an additional broadening parameter was needed to fit the data (see text).
} \label{Fig2}
\end{figure}

All the studied films were grown ex situ. No further treatment was performed prior to STM/STS experiment. Constant current STM images revealed a very smooth film surface with $\approx0.1$~nm roughness, which is 10 times less than the corrugation of patterned films \cite{Semenov}. The STS was performed at variable temperatures (2.5-20~K) using mechanically cut Pt/Ir tips. The local tunneling conductance spectra $dI/dV(V,x,y)$ were obtained by numerical derivative of raw $I(V)$ data. They are directly linked to the local quasiparticle excitation spectrum, $N_{S}\,(E, x, y)$, through the relation $dI(V)/dV \propto - \int\ dE N_{S}\,(E+eV,x,y)\partial f(E)/\partial E$, where $f(E)$ is the Fermi-Dirac function.

The thicker film, 15~nm, shows conventional and \emph{spatially homogeneous} SC conductance spectra, with a non-broadened BCS-like shape, characterized by a gap $\Delta (0)= 2.85$~meV (Fig.1); the gap was observed to vanish at the film's $T_C$ extracted from $R_{sq}(T)$ measurement. Upon application of a magnetic field, a well-known disordered vortex lattice was revealed, attesting for local disorder existing in this extreme type II SC (Fig.1d). Thus, our STS data on the 15~nm sample are in good agreement with the conclusions of previous reports \cite{Kirtley,Kashiwaya_PhysicaB,VanBaarle,Chockalingam}. %Thin NbN films were also subject of several STM/STS studies in the past. Kirtley et al. \cite{Kirtley} observed the tunneling spectra of conventional BCS-like shape and reported a SC gap $\Delta (0)\approx 2.5$~meV. The conventional SC state in NbN was confirmed by Kashiwaya et al. \cite{Kashiwaya_IEEE}; they also studied the magnetic field response \cite{Kashiwaya_PhysicaB} and extracted the effective coherence length of approximately 7~nm from the vortex core size. More recently, STM/STS imaging of the vortex structures in $\approx$ 60~nm thick NbN film (covered by a thin Au layer) was performed by Nishizaki et al. \cite{Nishizaki,VanBaarle} while Chockalingam et al. \cite{Chockalingam} studied $\approx50$~nm thick one; both groups observed a disordered Abrikosov lattice. None of these papers aimed addressing the SIT problem.

Figure~2 summarizes our STS results obtained on thinner films (8, 4, 2.33~nm). The data are organized in three columns by thickness. From top to bottom, each column shows: i) a normalized conductance map for a bias voltage chosen to picture out possible gap inhomogeneities, ii) selected spectra representative of the maximal variations observed among the local $dI/dV(V)$ spectra (red and blue curves) and typical spectrum in magnetic field at vortex centers (black curve), iii) an image of the vortex lattice and iv) the temperature evolution of local spectra.

The 8~nm $dI/dV$ map reveals only tiny spatial inhomogeneities of the SC gap, except in regions where structural defects occur, such as the red patch in the image. The disordered vortex lattice is routinely observed. The temperature evolution of the tunneling spectra remains conventional, although a slight spectral broadening was detected and accounted for with a very small Dynes pair-breaking parameter $\Gamma \approx 0.01$~meV. In overall, the 8~nm NbN film behaves similarly to the 15~nm one.

The situation changes in the 4~nm sample. While the vortex lattice and the temperature evolution of the tunneling spectra do not manifest significant differences with respect to thicker samples, small spatial inhomogeneities are revealed in the tunneling conductance map, which
specifically affect the gap width and the height of the coherence peaks (see the red and blue curves below the conductance map). These inhomogeneities are characterized by two spatial length scales. The smaller one is of the order of a few nm, thus close to the coherence length scale. The larger one is of several tens of nm; further analysis showed that the latter originates from slight variations of the film thickness due to underlying atomic steps of the substrate (see also Fig.~3 and section 3 of the supplemental material). Nevertheless, the $dI/dV$  spectra at $T \ll T_c$ remain reasonably BCS-like, the fit again requiring $\Gamma \approx 0.01$~meV.

In the 2.33~nm sample, the tunneling spectra undergo more and more pronounced changes. The conductance maps reveal inhomogeneities comparable to the 4~nm case, characterized by two length scales. "Full" gap in $dI/dV$ exists everywhere, characterized by zero conductance and coherence peaks, with however their amplitude much reduced with respect to the case of thicker films; their height is observed to vary spatially at the nanometer scale (see the blue and red curves), similarly to the effects discovered in cuprates \cite{Cren_2001}. The vortex lattice is hardly observable; Inside the cores the spectra present a large dip with no coherence peaks. Moreover, the normal state spectra reveal a V-shaped background with a clear minimum at zero-bias, extended on a larger energy scale than the SC gap. A shallow pseudogap (a dip opening above $T_C$) is also observed within a 1~K range above $T_C$, whose energy scale is comparable to the SC gap. The spectra are poorly described by a BCS expression at $T \ll T_c$ and require $\Gamma \approx 0.05$~meV (see fit).
%As the film transits to the SC phase, the SC gap opens on this voltage dependent spectral background (black curves in Fig.2). At first sight, as the temperature is lowered and the SC gap becomes more and more pronounced, it seems that the "V"-shaped background reduces smoothly and almost disappears, thus "restoring" at low temperature a BCS-kind tunneling spectrum with an almost flat background (see $dI/dV$ curve at the very bottom of Fig.2). However, the coherence peaks appear strongly dumped when compared to the case of thicker samples.

%\begin{figure}
%\begin{center}
%\includegraphics[width=7.5cm]{Fig3.eps}
%\end{center}
\begin{figure}[h]
\hspace{-0.3 cm} \epsfxsize = 8.7 cm \epsfbox{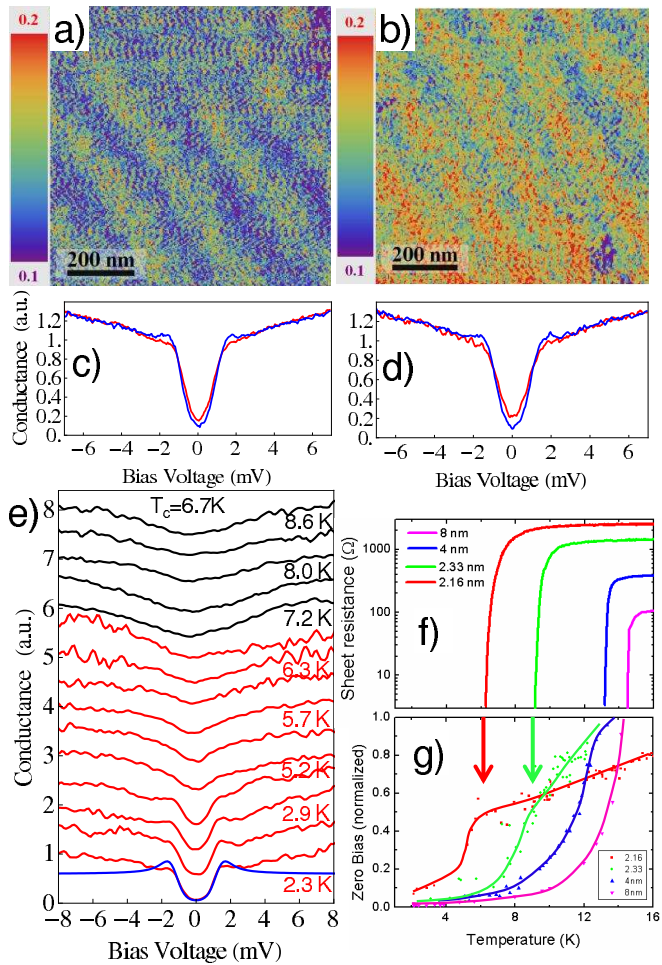}
\caption{(Color online)
Tunneling characteristics of the 2.16~nm thick film at 2.3~K. a) zero-bias STS conductance maps at $B=0$~T, b) at $B=3$~T. c) Typical variations among the corresponding local spectra for $B=0$~T and d) for $B=3$~T; e) Evolution of the $dI/dV(V)$ spectra with temperature. The bottom thick blue line is a BCS calculation at 2.3~K with an additional broadening parameter (see text). f) zoom on $R(T)$ of the 2.16-8~nm films compared to their g): temperature dependence of the zero-bias conductance $dI(V=0)/dV$.
} \label{Fig3}
\end{figure}

Striking changes are revealed in the tunneling characteristics of the 2.16~nm sample (see Fig.~3). The SC gap develops on a much deeper V-shaped background, existing also above $T_C$ (Fig.~3e) thus characteristic of the normal state, the energy scale of which is of 5-10 meV, similar to the one observed for the 2.33~nm sample. In addition, a very shallow pseudogap is present; it appears more clearly in Fig.~3f-g where $R_{sq}(T)$ and $dI/dV(V=0,T)$ are presented simultaneously for 2.16-8~nm films. The difference between the two behaviors is straightforward : In thicker samples, the SC gap opens at $T_C$ on a flat spectral background (usual metallic state), while in the thinner 2.16 and 2.33~nm samples, the DOS at $E_F$ is already reduced in the normal state. Additionally, the opening of the SC gap occurs at $T_C$, as it is unambiguously detected as a kink in the $dI/dV(V=0,T)$ curves (pointed by the arrows), in contrast to the smoother evolution in high-$T_C$ cuprates (see \cite{Fischer_Review} and ref. therein).

Furthermore, the spectra are not BCS-like anymore; using nevertheless a broadened BCS DOS to estimate the gap value, yields $\Gamma \approx 0.10$~meV. The amplitude of the quasiparticle peaks is further damped with respect to the 2.33~nm case, so that in some areas the peaks disappear completely. The conductance map in Fig.~3a allows again visualizing two-scale spatial variations, as observed in 4 nm and 2.33 nm samples, with spectral inhomogeneities quite comparable to these latter cases. i) The zero-bias conductance and the height of quasiparticle peaks vary slightly at the nm scale. ii) The large scale variations appear strikingly to be spatially ordered in parallel bands in some areas of the sample (Fig.~3a-b; see section 3 of the supplemental material for more details);

Another astonishing feature is observed : the spectra are dramatically unsensitive to the magnetic field (Fig.~3b-d); up to B=3~T, no vortices could be observed (Fig.~3b). However, while bulk $H_{C2}$ is estimated to be $\approx 20$~T, $H_{C2}$ decreases with film thickness, but remains larger than 10~T for all our studied films and is thus much higher than the fields used here. Vortices are a mesoscopic manifestation of the quantum coherence of the SC condensate, where the phase of the order parameter is spatially well-defined and makes $2\pi$ turns around each vortex core. Thus, we interpret this striking result as a break-up of the long-range phase coherence in this fragile SC condensate, suggesting a lowered phase stiffness already at $T=0.4T_{C-bulk}$, thus quite before the very SIT.

%\begin{figure}
%\begin{center}
%\includegraphics[width=7.5cm]{Tab1.eps}
%\end{center}
%\begin{figure}[h]
%\hspace{-0.3 cm} \epsfxsize = 5.0 cm \epsfbox{FigTab1.eps}
%\caption{} \label{Tab1}
%\end{figure}

\begin{table}%[H] add [H] placement to break table across pages
\caption{\label{GapTcvalues} Critical temperature $T_C$ (extracted from transport measurements), energy gap value
 $\Delta_{BCS}$ (extracted from BCS fits of the tunnel conductance spectra) and $2\Delta_{BCS}/k_BT_C$ ratio as a function of film thickness $d$. Although the tunnel conductance spectra for the thinnest films at the lowest studied temperature (2.3 K) deviate from BCS behavior, we estimate the energy gap from a broadened BCS simulation.}
\begin{ruledtabular}
\begin{tabular}{c|lllll}
$d$(nm) & 15 & 8 & 4 & 2.33 & 2.16 \\
$T_C$(K) & 15.0 & 14.5 & 13.3 & 9.4 & 6.7 \\
$\Delta_{BCS}$ & 2.85 & 2.7 & 2.4 & 1.7 & 1.3 \\
2$\Delta_{BCS}$/$k_B$$T_C$ & 4.4 & 4.3 & 4.2 & 4.2 & 4.0 \\
\end{tabular}
\end{ruledtabular}
\end{table}

A hallmark of our tunneling data on approaching the SIT is the development of a more and more pronounced V-shaped background when the film thickness is reduced. This suggest enhanced Coulombic effects \cite{Altshuler} and is consistent with the fact that our $T_C$ vs $R_{sq}$ data follow the Finkelstein law \cite{note_supplmat} (see figure and section 1 of the supplemental material). The observed $\Delta$ and $T_C$ progressive reduction (see Tab.~\ref{GapTcvalues}), with smoothly decreasing $2\Delta_{BCS}/kT_C$ ratio, is consistent with previous reports on homogeneous disordered thin films close to the SIT \cite{Goldman} and with the behavior found for much thicker NbN films by changing the disorder \cite{Chockalingam}. However, it is at variance with recent STS reports on TiN and InO films, both suggesting a huge increase of this ratio with increasing disorder \cite{Sacepe_TiN,Sacepe_InOx}.

It is instructive to make a detailed comparison of our local results on NbN with the recently reported STM/STS experiments on TiN (3.6-5~nm), InO (15 and 30~nm) and thicker (50~nm) NbN films.

The most important difference with respect to TiN and InO samples, is that a very deep pseudogap above $T_C$ was reported, where at $T_C$, almost all the DOS is depressed at $E_F$ \cite{Sacepe_TiN,Sacepe_InOx}. Additionally, it is important to note that these systems were closer to the SIT ($T_C \le 0.4 T_{C-bulk}$). In our case, all $T_C \ge 0.4 T_{C-bulk}$ and the largest depressed DOS at $E_F$, obtained for the closest films to the SIT, is of $35-50\%$, but is mostly due to the V-shaped background and not to the pseudogap effect. On the other hand, there are also similarities: Observation of small gap inhomogeneities and suppression of the coherence peaks with increasing disorder while the conductance spectra deviate more and more from BCS behavior.

The evolution of our tunneling spectra with varying thickness present many surprising similarities with recent studies on thicker NbN films, where the thickness was fixed and the SIT was approached by tuning the disorder (from $k_Fl \sim 1.2-10.1$) \cite{Chockalingam}. In particular, thicker NbN films present a $T_C$ versus $k_Fl$ dependence quite comparable to ours, and lead to tunneling characteristics similar to what is reported here. In particular, the features reported for $T_C=6$~K ($k_Fl =1.6$) and $T_C=4.1$~K ($k_Fl =2.2$) are comparable with those presented for our 2.16~nm film ($T_C=6.7$, $k_Fl \leq 2.6$) in terms of: i) a large broadening of the spectra leading to gap filling, ii) almost vanishing of the coherence peaks, iii) the existence of a shallow pseudogap above $T_C$.

%For TiN and InO it was suggested that the pseudogap and the SC gap energy scales are comparable (the $dI/dV$ spectra are shown only for $\left|V\right| \leq 1.5$~mV). However our results clearly show that in NbN the V-shaped background is characterized by a larger energy scale (5-10~mV) than the SC gap. The latter feature together with the observed kinks in $dI/dV(V=0,T)$ characteristics provide strong evidence for a separate electronic nature of these two spectroscopic features. %While in \cite{Sacepe_TiN} it was first concluded that the results were consistent with various theoretical models, it was further argued that superconducting fluctuations, which effect was fitted in a small temperature range ($1.1-2.7~T_c$ ), could explain the pseudogap state. However this pseudogap was observed in a much larger temperature range (up to $14~T_c$ for the most disordered sample) which suggests other possible mechanisms. To explain further on the InO results and to reinterpret the TiN ones, another mechanism based on localization of preformed Cooper pairs was proposed, without however a direct proof of the existence of such preformed pairs.

For InO films, it was argued in \cite{Sacepe_InOx} that the disappearance of the coherence peaks occurs at $T_C$. This is a noticeable difference with our results, where the decay of the coherence peaks amplitude is observed to be gradual as a function of thickness and $T_C$ reduction. The coherence peaks amplitude almost vanishes in the SC state (thus below $T_C$) when $T_C \approx 0.4T_{C-bulk}$, together with the concomitant disappearance of the vortex lattice, indicating the break-up of long range phase coherence of the SC condensate. Parts of the observed significant differences between NbN and TiN/InO films may be related to a more homogeneous crystalline structure of the NbN system, which is preserved and controlled upon thickness reduction (the growth conditions being identical for all films), leading to a more continuous behavior of the electronic properties toward the SIT.%, being possibly compatible with granular superconductivity scenarios as suggested by former results on InO films \cite{Hebard}.

In conclusion, the present work shows that already at $T_C \approx 0.4T_{C-bulk}$, thus quite before the SIT, phase fluctuations of the SC order parameter lead to the almost vanishing of the coherence peaks, together with the disappearence of the vortex lattice, while the SC state at zero magnetic field show only slight spatial spectral inhomogeneities. Therefore, numerical simulations of a strongly disordered \emph{s}-wave superconductor, showing that amplitude fluctuations alone are not sufficient to drive the SIT \cite{Ghosal} are compatible with our results and with more recent numerical study \cite{Bouadim}. Further investigations very close to the SIT are necessary to clarify the role and the nature of the pseudogap and of the V-shaped background we observed in ultrathin homogeneous NbN films.

The authors acknowledge fruitful discussions with B. Sac\'ep\'e. The work is supported in part by DFG Center for Functional Nanostructures under sub-project A4.3.

\end{document}